# Versatile sputtering technology for Al$_2$O$_3$ gate insulators on graphene

M. Friedemann[1], M. Woszczyna[1], A. Müller[1], S. Wundrack[1], T. Dziomba[1], Th. Weimann[1], Th. Seyller[2], F. Ahlers[1,*]

[1] *Physikalisch-Technische Bundesanstalt, Bundesallee 100, D-38116 Braunschweig, Germany*
[2] *Lehrstuhl für Technische Physik, Universität Erlangen-Nürnberg, 91058 Erlangen, Germany*
[*] *Author to whom correspondence should be addressed. Electronic mail: franz.ahlers@ptb.de.*

We report a novel fabrication method of graphene Al$_2$O$_3$ gate insulators based on sputtering. Electrical performance of dual-gated mono- and bilayer exfoliated graphene devices is presented. Sputtered Al$_2$O$_3$ layers possess comparable quality to oxides obtained by atomic layer deposition (ALD) with respect to a high relative dielectric constant of about 8, as well as low-hysteresis performance and high breakdown voltage. We observe a moderate carrier mobility of about 1000 cm$^2$/Vs in graphene and 350 cm$^2$/Vs in its bilayer due to increased resonant scattering on atomic scale defects. Most likely this originated from the thin Al precursor layer evaporated prior to sputtering the Al$_2$O$_3$ gate oxide.

Graphene is a single atom thick layer of carbon. One of its most important properties is the relatively easy charge carrier tuneability, typically realized by an electrostatic gate.[1] Due to this and to its high mobility, graphene is predicted to be the future material in semiconductor industry[2] and quantum resistance metrology.[3] To date, it is extensively studied both theoretically and experimentally in fundamental science. Since the beginning of exfoliated graphene research thermally oxidized silicon is successfully used as both a graphene flake support and as a back-gate insulator[4]. However, a great demand for alternative technologies appeared when it became clear that epitaxially grown graphene on SiC may fulfill industrial requirements.[5] Moreover, in order to explore effects occurring in p-n junctions (PNJ)[6,7], electron waveguides[8], or single electron transistors,[9] high quality dielectric material is required to form efficient gates for these devices.

To date, the major efforts of graphene top-gate fabrication are focused on utilizing HfO$_2$ and Al$_2$O$_3$ as high-κ materials in atomic layer deposition (ALD) facilities.[10,11] ALD may provide a high quality dielectric with a precisely defined thickness. On the other hand, due to the hydrophobic behavior of graphene, ALD may suffer from non-uniform material deposition.[10] To overcome this, an additional technological step is introduced, in which the graphene surface is pretreated before main dielectric fabrication.[11] Recently, it has been shown that deposition and subsequent oxidation of 1-2 nm of aluminum sufficiently prepares the graphene area before the ALD process and leads to a more uniform Al$_2$O$_3$ dielectric coverage.[11-13] In our work we employed the same approach, but instead of utilizing ALD we used sputtering, which is much simpler and economically more favorable than ALD. To our knowledge, we present in this paper the first transport results of graphene top-gated by sputtering technology.

Before fabrication of top-gates graphene samples were prepared in a standard procedure. Graphene flakes from natural graphite were exfoliated onto Si/SiO$_2$ and, using e-beam lithography, they were then shaped into a Hall bar geometry by Ar/O$_2$-ion etching. Finally Ti/Au contacts (10 nm/50 nm) were evaporated. In the next process phase, in order to fabricate top-gates, samples were again coated by PMMA, and masks for Al$_2$O$_3$ layers were made by e-beam lithography. Next 2 nm of aluminum was thermally evaporated at a rate of 0.1 nm/s. The aluminum fully oxidized after removing the samples from the chamber and acted as a protection layer during the subsequent Al$_2$O$_3$ oxide deposition. In this step a 26 nm thick Al$_2$O$_3$ layer was sputtered on the samples by an RF sputtering process from a pure Al$_2$O$_3$ target using a power of 350 W. After sputtering the Al$_2$O$_3$ layer the samples were heated for 10 minutes at 400°C in order to anneal the dielectric internal structure. An acetone lift-off was performed during 12 hours to define the final dielectric shape. In the last fabrication phase, top-gate electrodes were created in the same fashion as the graphene contacts. In the described approach the dielectric area was made larger than the graphene sheet and the top-gate electrode (see Fig. 1(h)) to ensure that the graphene was completely covered and that there was no direct connection between graphene and the metal gate. Since the thickness of the oxide is smaller than the thickness of the contact metallization, it may happen that the oxide layer does not perfectly cover the contacts and that discontinuities of the dielectric film appear at contact metallization edges. In previous experiments with evaporated HfO$_2$ of 80 nm thickness we had often found cracks in the oxide at the contact metallization boundaries and that upon evaporating the gate metal on the damaged areas, low-resistance paths between the electrical contacts and the top-gate metallization had been produced. To prevent this, we left a small lateral gap (~500 nm) between the top-gate metallization and the contact area. We kept this layout for all our Al$_2$O$_3$ samples. Further reduction of this safety distance should be possible by optimizing the lithography process. In addition, we successfully deposited a conformal dielectric layer onto a SiC substrate without graphene using the same Al$_2$O$_3$ oxide technology, as a preliminary experiment before fabrication of top-gates on epitaxial graphene on SiC.

In order to locally investigate the quality of the dielectric layers, atomic force microscopy (AFM) imaging was employed using a Nanostation II AFM by SIS (now N8 Neos by Bruker) in intermittent contact mode (tapping-mode). Apart from the topography image, we have simultaneously recorded the phase shift map of probe vibrations (the phase shift between the mechanical excitation of the cantilever base and the resulting mechanical oscillation of the cantilever tip), which allows to reveal local material inhomogeneities on the surface. Fig. 1 (a) and (b) show an AFM image and a line profile of $Al_2O_3$ deposited on exfoliated graphene, whereas Fig. 1 (d) and (e) depict an AFM image and a profile of $Al_2O_3$ on the SiC substrate. For the latter we omitted the annealing step at 400°C. All investigated samples exhibited continuous oxide layers with a mean *rms* surface roughness of about 0.6 nm. For annealed samples we observed in both topography and phase shift images distinctive but sparse features of maximum lateral size of approximately $20 \times 20$ nm$^2$ (Fig. 1(c)). Such features were not observed in not annealed samples, and we therefore conclude that the annealing procedure introduced contaminations probably coming from the environment and not from the dielectric itself, since the phase shift images revealed high composition contrasts in those areas.

To assess the influence of the top-gate dielectric on graphene quality, we employed Raman spectroscopy. Raman spectra were obtained using a LabRAM ARAMIS micro-Raman spectrometer (Horiba Jobin Yvon) equipped with a thermoelectrically cooled CCD detector (-70°C) and 100× objective (NA 0.90). The graphene samples were excited by a frequency-doubled Nd:YAG laser with at 532 nm wavelength. Fig. 1(f) shows a comparison of Raman spectra of exfoliated graphene on $SiO_2$ covered by $Al_2O_3$ (red trace) and of an uncovered graphene flake on $SiO_2$ (black trace). The Raman spectrum of $Al_2O_3$ covered graphene is dominated by a D–peak at ~1345 cm$^{-1}$ arising from a ouble-resonance Raman process, indicative of electron scattering on graphene defects.[14] This is also attested by the occurrence of a small D'–peak at ~1625 cm$^{-1}$.[15] Consequently, the D– and D'–peaks indicate an increased number of lattice defects in the $Al_2O_3$ covered graphene. Furthermore, in comparison to the as-exfoliated graphene, shifts of the G–peak (~11 cm$^{-1}$) and the 2D-peak (~18 cm$^{-1}$) were clearly observed. These significant shifts may have been generated during the annealing process by enhanced adhesion interactions between graphene and dielectric layers, which induce strain forces into the graphene lattice.[16]

Transport measurements of all fabricated devices were carried out in four-terminal configuration in vacuum of approximately $10^{-6}$ mbar at room temperature, and some of the samples were also characterized at low temperature of 1.6 K. Fig. 1(g) presents a typical graphene device resistivity dependence on both top-gate and back-gate voltages. The respective counter-gate was kept at the ground potential. The experimental setup is schematically depicted in inset (h) of Fig.1. In order to extract quantitative information about the electronic properties, measurement data have been fitted by the expression $\rho(n) = 1/\mu e n + \rho_s$, where $\mu$ is the field effect mobility, $e$ the elementary charge, $n$ the graphene carrier concentration, and $\rho_s$ the residual resistivity.[18] The carrier concentration was estimated based on the formula $n^2 = n_0^2 + [C_G(V_G - V_{G0})/e]^2$, where $n_0$ is a residual carrier concentration, $C_G$ is a gate capacitance per unit area, and $V_G$ and $V_{G0}$ are gate voltage and voltage at charge neutrality point (CNP), respectively. We ignored a quantum gate capacitance contribution in our model.[12] As is seen in Fig. 1(g) we obtained excellent fits to the experimental data. Residual carrier concentrations for top- and back-gate curves were similar, about $5.5 \times 10^{11}$ cm$^{-2}$ and $5.9 \times 10^{11}$ cm$^{-2}$, respectively. Also residual resistivity values were similar, about 1.4 kΩ in both cases. The mobility extracted from top-gate dependent resistivity was about 10% higher than from back-gate measurements, the values reaching 1100 cm$^2$/Vs and 1000 cm$^2$/Vs, respectively. While the residual carrier concentrations of top-gated samples were practically the same as those of samples fabricated in the same fashion but without top-gate, we observed a severe drop of carrier mobility and a dramatic raise of residual resistivity in the former. Samples without top-gates typically showed a mobility of 3000–5000 cm$^2$/Vs and 1000–1500 cm$^2$/Vs for graphene and bilayer graphene, respectively. The residual resistivity typically was below 300 Ω. Al$_2$O$_3$ covered structures, however, had mobilities around 1000 cm$^2$/Vs for graphene and 350 cm$^2$/Vs for its bilayer, while residual resistivity was in the range of 1.3-1.6 kΩ. These observations are in line with the existence of the D- and the D'-peaks in the Raman spectra, since measured $\rho_s$ values are independent of graphene carrier concentration and can be attributed to short-range, resonant carrier scattering on lattice defects.[17,18] It has been also shown that resonant scatterers are responsible for limiting graphene mobility, which would explain the low mobility in our top-gated devices.[19] A further improvement of the fabrication process of Al$_2$O$_3$ top-gates for applications requiring high carrier mobility like transistors or quantum resistance standards might be achieved by covering graphene sheets with a thin buffer layer before oxide sputtering. This additional layer would protect graphene against atomic scale deformations during dielectric deposition.[20] However, in applications focusing on other graphene electrical properties, like for instance graphene memory devices[21] this auxiliary fabrication step may be skipped. Furhter, it has been reported that already the Al-precursor top-coating may severely deteriorate the mobility.[22,23] This suggests that one may also optimize the formation step of this layer, since it is the one in the direct contact with the graphene sheet. The subsequent Al$_2$O$_3$ sputtering procedure should have a minor contribution to the graphene mobility drop.

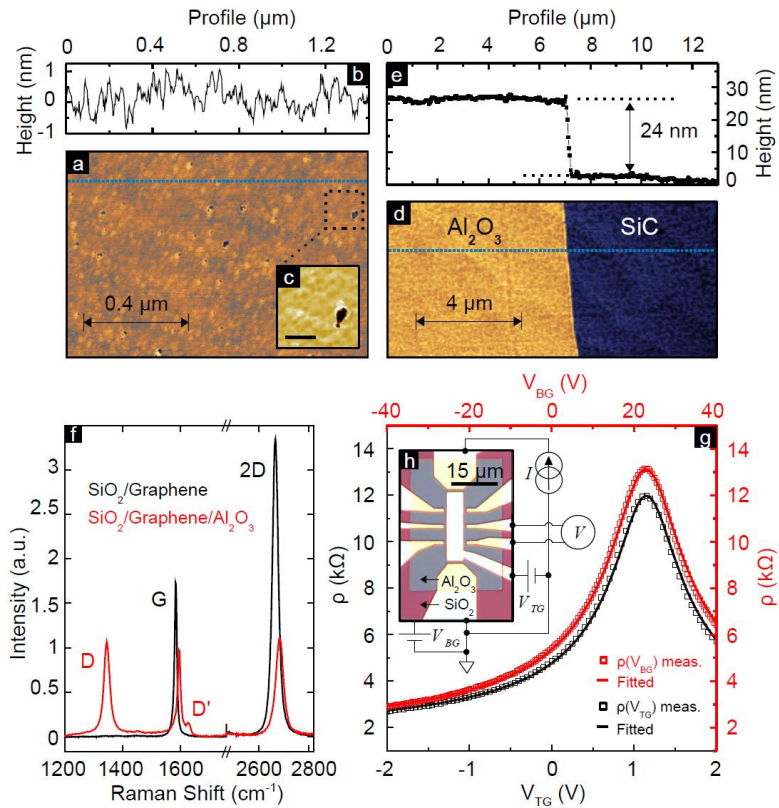

Fig. 1. (Color online) (a) AFM surface imaging of $Al_2O_3$ dielectric. (b) A typical profile of the surface taken along the dashed line in (a). Surface *rms* roughness of fabricated $Al_2O_3$ layers was about 0.6 nm. Inset (c) in (a) depicts a phase shift of an AFM probe signal, which reveals a point material contrast in the area marked in (a). The inset scale bar indicates 50 nm. (d) AFM surface imaging of a homogenous $Al_2O_3$ layer deposited on SiC. No morphological defects in the dielectric structure have been found. (e) The profile taken for the dashed line in (d). (f) Raman spectroscopy of a graphene device on $SiO_2$ substrate without (black trace) and with (red trace) $Al_2O_3$ dielectric on top made by sputtering. (g) Graphene resistivity in a function of applied top- and back-gate voltages (lower black and upper red traces, respectively). A counter-gate was kept at the common potential, while sweeping a gate voltage. Inset (h) presents a fabricated graphene Hall-bar and an electrical connection scheme.

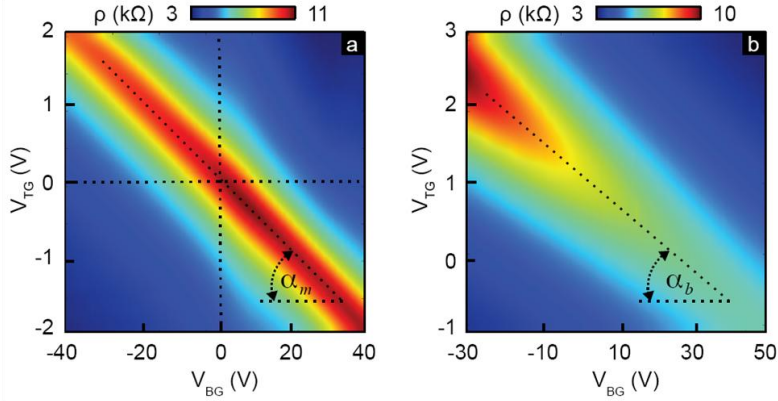

Fig. 2. (Color online) Mapping of graphene (a) and graphene bilayer (b) Transport measurements were carried out in four-terminal configuration at room temperature with a DC current of 1 µA.

Fig. 2(a) and (b) present graphene and bilayer graphene resistivity maps in dependence on top- and back-gate voltage. In the monolayer graphene device only a linear shift of the CNP is visible, whereas in the bilayer a pronounced raise of resistance at CNP for increasing opposite top- and back-gate voltages clearly indicates gap opening between valence and conduction bands.[24] In dual-gated measurements the carrier concentration formula is modified to $n^2 = n_0^2 + [C_{TG}(V_{TG} - V_{TG0})/e + C_{BG}(V_{BG} - V_{BG0})/e]^2$, where subscripts *TG* and *BG* denote top- and back-gate.[25] Based on the CNP slope $\tan(\alpha) = C_{TG}/C_{BG}$ (see diagonal dashed lines in Fig. 2(a) and(b)), which expresses the carrier tuning efficiency ratio between top- and back-gate, we were able to estimate $C_{TG}$. From the known thickness (300 nm) and dielectric constant (3.9) of the $SiO_2$ back-gate insulator we calculated $C_{BG}$ of $1.1 \times 10^{-4}$ F/m$^2$ and thus obtained, from the measured slopes $\tan(\alpha_m) = 20$ and $\tan(\alpha_b) = 23$, $C_{TG} = 22 \times 10^{-4}$ F/m$^2$ and $C_{TG} = 25.3 \times 10^{-4}$ F/m$^2$ for graphene and bilayer devices, respectively. Based on the AFM determination of oxide thickness of about 28 nm we then estimated an effective $Al_2O_3$ relative dielectric constant of 7.3 for the graphene device shown in Fig. 2(a) and 8.4 for the bilayer device shown in Fig. 2(b). These values prove the very high quality of the $Al_2O_3$ dielectric layers in our devices, which is comparable to the best results reported in literature for ALD-made $Al_2O_3$, where effective dielectric constants ranged from 6 to 8.[12,26,27]

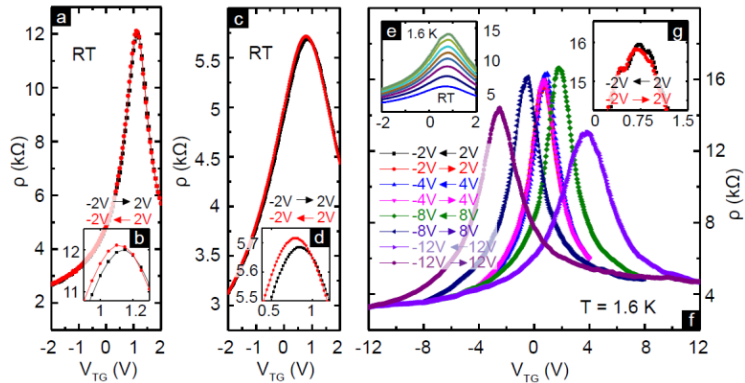

Fig. 3. (Color online) Hysteresis at room temperature in monolayer (a) and bilayer graphene (c) devices on SiO$_2$ covered by Al$_2$O$_3$. Insets (b) and (d) show magnified vicinities of CNPs in (a) and (c), respectively. (e) Bilayer resistivity changes for lowering temperatures from $T$ = 296 K to $T$ = 1.6 K. (f) Bilayer hysteresis development at T = 1.6 K. Top-gate voltage ranges were being increased successively from ± 2 V up to ± 12 V. Voltage sweep rate was 0.1 V/s. The inset (g) depicts the magnified vicinity of CNP for the top-gate voltage range ± 2 V for comparison to (b) and (d).

An imperfect structure of the dielectric layer leads to charge transfer within the oxide bulk, but this also may happen in the oxide/graphene interface. It has a strong influence on graphene electronic transport, often causing resistance hysteresis. Therefore we studied the hysteretic behavior of top-gated samples at room temperature $T = 296$ K and at $T = 1.6$ K (Fig. 3) to obtain additional qualitative information about the sputtered oxide. Only negligible hysteresis less than 60 mV of top-gate voltage and ~1% of resistivity change at the CNP was observed when the top-gate voltage was kept in a range of ± 2 V at room temperature,,indicating a low density of interface states. At low temperature (Fig.3(e)) the low-hysteresis performance was maintained up to ± 4 V. Fig. 3(f) presents the bilayer resistivity in a function of top-gate voltage, where sweep ranges have been successively increased from ± 2 V up to ± 12 V, at a constant sweep rate of 0.1 V/s, and the back-gate voltage at zero potential. When the sweep range exceeded ± 4 V, the resistivity curve became asymmetric and a hysteresis appeared, reaching 6.3 V of top-gate voltage and 11% of resistivity amplitude at CNP for a ± 12 V sweep. In general, we attribute the hysteretic behavior to a charge trapping mechanism in the oxide layer. For negative top-gate voltage, after exceeding a threshold of about ± 4 V at low temperature, holes from the graphene bilayer become trapped which dopes the graphene sheet into opposite polarity. This in turn weakens the top-gate efficiency of inducing carriers, but also shifts the CNP towards lower voltages due to additional electron doping. Analogously, positive top-gate voltage above the threshold shifts the position of CNP to higher voltages. At room temperature for top-gate voltages below trapping threshold, however, the hysteresis had a slight opposite direction, which may happen when the electrostatic potential around the graphene is changed e.g. by changing dipole orientations in oxide layers due to an applied external electric field.[28] We typically observed $Al_2O_3$ dielectric breakdown voltages at room temperature of about ± 9 V, whereas at low temperature the safe top-gate voltage range extended up to about ± 19 V. This means that the breakdown in the 28 nm thick oxides occurred for an electric field of approximately 0.32 V/nm, a value comparable to that of $SiO_2$ at room temperature.[28]

In summary, we have presented the first successful demonstration of a simple $Al_2O_3$ sputtering process in graphene gating technology. In terms of low-hysteresis performance, high breakdown voltage, and high effective dielectric constant, the quality of the dielectric layers was comparable to the best results of an ALD process. We have also reported a deterioration of the graphene mobility due to increased resonant scattering on atomic scale defects, but there is, in our opinion, still room for technology improvements. In further experiments, and when high mobility is crucial, one may introduce a buffer dielectric layer between graphene and the sputtered oxide, optimize technology of the Al precursor layer, and/or optimize the oxide post-processing annealing procedure.